\documentclass{ws-mpla}
\usepackage[super]{cite}
\usepackage{graphicx}
\usepackage{amsmath,amssymb,amsfonts, bm}
\usepackage{dsfont}
\usepackage{epsfig}
\usepackage{slashed}
\usepackage{color}

\usepackage{commath}
\usepackage{calc}

\newcommand{\be}{\begin{equation}}
\newcommand{\ee}{\end{equation}}

\newcommand\nnm{\nonumber}

\begin{document}



\title{Right-right-left extension of the standard model}

\author{Gauhar Abbas}

\address{IFIC, Universitat de Val\`encia -- CSIC, Apt. Correus 22085, 
E-46071 Val\`encia, Spain.\\
Gauhar.Abbas@ific.uv.es}

\maketitle

\begin{abstract}
A right-right-left extension of the standard model (SM) is proposed.  In this model, the standard model gauge group $SU(2)_L \otimes U(1)_Y $ is extended to $SU(2)_L \otimes SU(2)_R \otimes SU(2)^{\prime}_R \otimes SU(2)^{\prime}_L \otimes U(1)_{Y}$.  The gauge symmetries $SU(2)^{\prime}_R$, $SU(2)^{\prime}_L$ are the mirror counter-parts of the $SU(2)_L$ and $SU(2)_R$, respectively. Parity is spontaneously broken when the scalar Higgs fields acquire vacuum expectation values (VEVs) in a certain pattern.   Parity is restored at the scale of $SU(2)^{\prime}_L$.  The gauge sector has a unique pattern.  The scalar sector of the model is optimum, elegant and unique.  

\keywords{extension of the gauge sector; parity.}
\end{abstract}

\ccode{PACS Nos.: 12.60.Cn,12.60.Fr}

Left-right symmetric (LRS) models are one of the most attractive and aesthetics possibilities beyond the standard model (SM) \cite{Pati:1974yy,Mohapatra:1974hk,Senjanovic:1975rk,Senjanovic:1978ev}.   Parity is spontaneously broken in these models which leads to a rich phenomenology of new physics.  Furthermore,   LRS models have a natural explanation for neutrino masses and mixing.

In this work, we present an interesting right-right-left  extension of the SM with mirror symmetries. The idea of mirror symmetries was first discussed in Refs.\refcite{Lee:1956qn,Kobzarev:1966qya,Pavsic:1974rq}.  We shall see that predictions of the model presented in this work are within the reach of the high luminosity phase of the large hadron collider (LHC).  The starting point is to extend the SM gauge group $SU(2)_L \otimes U(1)_Y $ to $SU(2)_L \otimes SU(2)_R \otimes SU(2)^{\prime}_R \otimes SU(2)^{\prime}_L \otimes U(1)_{Y}$ where the symmetries $SU(2)^{\prime}_R$ and  $SU(2)^{\prime}_L$ are the mirror counter-parts of the gauge groups $SU(2)_L$ and $SU(2)_R$, respectively.  The left- and right-handed fermions on the SM side, live in the fundamental representation of the gauge groups $SU(2)_L$  and $SU(2)_R$.  This is similar to LRS models\cite{Pati:1974yy,Mohapatra:1974hk,Senjanovic:1975rk,Senjanovic:1978ev}.  On the other side, the mirror fermions are accommodated in the mirror-counter parts of the $SU(2)_L$  and $SU(2)_R$.  This is a unique feature of the model which is not explored in the literature yet. 

Under parity, fermionic fields of the theory are assumed to transform as following:

\be
 \psi_L  \longleftrightarrow \psi^{\prime}_R,~ \psi_R \longleftrightarrow \psi^{\prime}_L, 
\ee
where, $\psi_L$, $\psi_R$ are doublets of the  symmetries $SU(2)_L$ and  $SU(2)_R$, respectively.  The doublets $\psi^{\prime}_L$, $\psi^{\prime}_R$ correspond to the gauge groups $SU(2)^{\prime}_L$ and  $SU(2)^{\prime}_R$.  The doublets $\psi_L$, $\psi_R$ are assumed to be singlets under the gauge groups $SU(2)^{\prime}_L$ and  $SU(2)^{\prime}_R$.   The doublets $\psi^{\prime}_L$, $\psi^{\prime}_R$ are singlet under  the gauge symmetries  $SU(2)_L$ and  $SU(2)_R$.  

The  gauge fields corresponding to the non-Abelian gauge symmetries of the theory transform exactly in the same manner under parity.  Thus, the behavior of the gauge fields under parity is given by
\be
\mathcal{W} \longleftrightarrow  \mathcal{X^\prime},~  \mathcal{X} \longleftrightarrow  \mathcal{W^\prime},~  \mathcal{B}_\mu \longleftrightarrow  \mathcal{B}_\mu,
\ee
where $ \mathcal{W} $ and $ \mathcal{X} $ are the gauge fields corresponding to $SU(2)_L$ and $SU(2)_R$.  The gauge fields $ \mathcal{W^\prime} $ and $ \mathcal{X^\prime}$ correspond to the symmetries $SU(2)^{\prime}_L$ and $SU(2)^{\prime}_R$, respectively. The gauge field $\mathcal{B}_\mu $ correspond to the gauge symmetry group $U(1)_Y$. 

The hypercharge operator $Y$ can be written in the following way:

\be
\label{Yp}
Y = 2 (Q-T_{3L}-T_{3R}-T^{\prime}_{3R}-T^{\prime}_{3L}), 
\ee

where $T_{3L}$ and $T_{3R}$ are the generators of $SU(2)_L$ and $SU(2)_R$, and $T^{\prime}_{3L}$, $T^\prime_{3R}$ are the generators of $SU(2)^{\prime}_L$ and $SU(2)^{\prime}_R$, respectively.  The electromagnetic charge operator is denoted by $Q$.

The left- and right-handed fermions under the $SU(3)_c \otimes SU(2)_L \otimes SU(2)_R \otimes SU(2)^{\prime}_R \otimes SU(2)^{\prime}_L \otimes U(1)_{Y}$ symmetry transform as following:

\begin{eqnarray}
\label{Q} 
Q_{iL}&=&\begin{pmatrix}
  u \\
   d 
 \end{pmatrix}_{iL}:(3,2,1,1,1,\frac{1}{3}),~Q_{iR}= \begin{pmatrix}
  u \\
   d
 \end{pmatrix}_{iR}: (3,1,2,1,1,\frac{1}{3}),
\nonumber\\
\label{L} L_{iL}&=& \begin{pmatrix}
  \nu \\
   l 
 \end{pmatrix}_{iL}:(1,2,1,1,1,-1),L_{iR}=\begin{pmatrix}
  \nu \\
   l
 \end{pmatrix}_{iR} :(1,1,2,1,1,-1)\nnm \\
 &
\end{eqnarray}

and

\begin{eqnarray}
\label{Q1} 
Q^{\prime}_{iL}&=&\begin{pmatrix}
  u^\prime \\
   d^\prime 
 \end{pmatrix}_{iL}:(3,1,1,1,2,\frac{1}{3}),~Q^{\prime}_{iR}= \begin{pmatrix}
  u^\prime \\
   d^\prime
 \end{pmatrix}_{iR}: (3,1,1,2,1,\frac{1}{3}),
\nonumber\\
\label{L1} L^{\prime}_{iL}&=& \begin{pmatrix}
  \nu^\prime \\
   l^\prime 
 \end{pmatrix}_{iL}:(1,1,1,1,2,-1),L^{\prime}_{iR}=\begin{pmatrix}
  \nu^\prime \\
   l^\prime
 \end{pmatrix}_{iR} :(1,1,1,2,1,-1),\nnm \\
 &
\end{eqnarray}

where $i=1,2,3$ correspond to number of generations and $Q_{iL,R}$ and $Q^{\prime}_{iL,R}$  denote the quark doublets, and $L_{iL,R}$ and$L^{\prime}_{iL,R}$ denote the leptonic doublets.  The quantum numbers for $SU(3)_c \otimes SU(2)_L \otimes SU(2)_R \otimes SU(2)^{\prime}_R \otimes SU(2)^{\prime}_L \otimes U(1)_{Y}$ are shown in parenthesis.   The quarks  $Q^{\prime}_L$ and $Q^{\prime}_R$, like their mirror counter-parts $Q_R$ and $Q_L$, also have colour charges under $SU(3)_c$ colour symmetry of the QCD.

The imposition of parity invariance constrains gauge couplings of the gauge groups.  Thus, the coupling $g_L$ of $SU(2)_L$ and $g^{\prime}_R$ of  $SU(2)^{\prime}_R$ are equal and given by $g_L = g^{\prime}_R=g_1$.  Similarly, the gauge coupling of $SU(2)_R$ and $SU(2)^{\prime}_L$ is $g_R = g^{\prime}_L=g_2$.  Therefore, covariant derivative reads

\begin{eqnarray}
 \mathcal{D}_\mu &\equiv & \partial_\mu -  i g_1 \left( \mathcal{W}^{i}_{\mu} T^{i}_L + \mathcal{X}^{\prime i}_{\mu} T^{\prime i}_R \right) -  i g_2 \left( \mathcal{X}^{i}_{\mu} T^{i}_R + \mathcal{W}^{\prime i}_{\mu} T^{\prime i}_L \right) - i g^\prime \mathcal{B}_\mu \frac{Y}{2}, 
\end{eqnarray}

where $T^{i}_{L,R}$ and $T^{\prime i}_{L,R}$ are the generators of $SU(2)_L$, $SU(2)_R$ and $SU(2)^{\prime}_L$, $SU(2)^{\prime}_R$, respectively.  The generator $Y$ corresponds to the gauge symmetry group $U(1)_Y$.  

Now we discuss patterns of the spontaneous symmetry breaking (SSB).  There are two ways to reach the SM group symmetry via SSB.  The first scenario is to break the whole symmetry $SU(2)_L \otimes SU(2)_R \otimes SU(2)^{\prime}_R \otimes SU(2)^{\prime}_L \otimes U(1)_{Y}$ to $SU(2)_L \otimes SU(2)_R \otimes  SU(2)^{\prime}_R \otimes U(1)_{Y^{\prime}}$ spontaneously.   In the  alternative way, the full symmetry is broken down to $SU(2)_L \otimes SU(2)_R \otimes SU(2)^{\prime}_L  \otimes U(1)_{Y^{\prime}}$.  The first scenario provides a symmetrical pattern of the scalar fields participating in the SSB and seems more elegant. Hence, we adopt this scenario in this work.

For initiating the SSB, we introduce two bi-doublets Higgs fields, $\varphi$ and $\chi$ for providing the masses to the SM and mirror fermions. In addition to this, we may introduce singlets, doublets or triplets to raise the scale  at every stage of the SSB.  The doublet and triplet representations are particularly interesting.  The triplet representation of the Higgs field is used in the so-called minimal left-right symmetric model (MLRSM) whose gauge symmetry is a subset of the symmetry considered in the present work \cite{Mohapatra:1979ia,Mohapatra:1980yp}.  The triplet representation leads to the Majorana mass of neutrinos.

Hence, the scalar sector of the model under the $SU(2)_L \otimes SU(2)_R \otimes SU(2)^{\prime}_R \otimes SU(2)^{\prime}_L \otimes U(1)_{Y}$ symmetry transforms as following:

\begin{eqnarray}
\chi= \begin{pmatrix}
  \chi^{0}_1 & \chi^{+}_1 \\
   \chi^{-}_2 & \chi^{0}_2 
 \end{pmatrix}:(1,1,2,2,0)
 \end{eqnarray}
and

\begin{eqnarray}
\Delta_L= \begin{pmatrix}
  \delta^{+}_L /\sqrt{2}& \delta^{++}_L \\
   \delta^{0}_L & -\delta^{+}_L/\sqrt{2}
 \end{pmatrix}: (1,1,1,3,2)
 \end{eqnarray}
 or
 
 \begin{eqnarray}
\chi_L= \begin{pmatrix}
  \chi^{+}_L \\
   \chi^{0}_L 
 \end{pmatrix}:(1,1,1,2,1).
 \end{eqnarray} 
 
The above configuration will cover the first two steps of the SSB.  The final two steps of the SSB which lead to the $U(1)_{EM}$ are performed by the following scalar fields:

\begin{eqnarray}
\varphi= \begin{pmatrix}
  \varphi^{0}_1 & \varphi^{+}_1 \\
   \varphi^{-}_2 & \varphi^{0}_2 
 \end{pmatrix}:(2,2,1,1,0)
 \end{eqnarray}

and
 
 \begin{eqnarray}
\Delta_R= \begin{pmatrix}
  \delta^{+}_R/\sqrt{2} & \delta^{++}_R \\
   \delta^{0}_R & -\delta^{+}_R/\sqrt{2}
 \end{pmatrix}: (1,3,1,1,2)
 \end{eqnarray}
 or
 
 \begin{eqnarray}
\varphi_L= \begin{pmatrix}
  \varphi^{+}_L \\
   \varphi^{0}_L 
 \end{pmatrix}:(1,2,1,1,1).
 \end{eqnarray} 

The SSB occurs when the neutral components of the scalar Higgs fields acquire a vacuum expectation value (VEV).   The  pattern of VEVs in the first two stages of the SSB is following:

\begin{eqnarray}
\langle \chi \rangle = \begin{pmatrix}
  v^{\prime}_1 & 0 \\
   0 & v^{\prime}_2 e^{i \alpha^\prime} 
 \end{pmatrix}
 \end{eqnarray}
and

\begin{eqnarray}
\langle \Delta_L \rangle = \begin{pmatrix}
  0 & 0 \\
   v_L e^{i \delta_L}  & 0 
 \end{pmatrix}~\textrm{or}~ \langle \chi_L \rangle = \begin{pmatrix}
  0 \\
    v_L 
 \end{pmatrix}.
 \end{eqnarray}

In the final two steps of the SSB, the VEVs of the scalar fields arrange themselves as given below

\begin{eqnarray}
\langle \varphi\rangle = \begin{pmatrix}
  v_1 & 0 \\
   0 & v_2 e^{i \alpha} 
 \end{pmatrix}
 \end{eqnarray}
and

\begin{eqnarray}
\langle \Delta_R \rangle = \begin{pmatrix}
  0 & 0 \\
   v_R e^{i \delta_R}  & 0
 \end{pmatrix}~\textrm{or}~\langle \varphi_R \rangle = \begin{pmatrix}
  0 \\
    v_R 
 \end{pmatrix}.
 \end{eqnarray}

Depending on whether triplet or doublet representations of the scalar Higgs fields are used, the pattern of the SSB can be described  in the following steps.

\begin{enumerate}
\item
The whole symmetry $SU(2)_L \otimes SU(2)_R \otimes SU(2)^{\prime}_R \otimes SU(2)^{\prime}_L \otimes U(1)_{Y}$  is broken down to $SU(2)_L \otimes SU(2)_R \otimes SU(2)^{\prime}_R \otimes  U(1)_{Y^{\prime}}$ when the scalar fields $\Delta_L$ or $\chi_L$ acquires a VEV.  The unbroken generator after this step is $Y^\prime = T^{\prime}_{3L} + \frac{Y}{2}$.

\item
After this, $SU(2)_L \otimes SU(2)_R \otimes SU(2)^{\prime}_R \otimes  U(1)_{Y^{\prime}}$ is broken down to $SU(2)_L \otimes SU(2)_R \otimes  U(1)_{Y^{\prime \prime}}$ by the VEV of the scalar field $\chi$. The  $Y^{\prime \prime} = T^{\prime}_{3R} +T^{\prime}_{3L} + \frac{Y}{2}$ is the unbroken generator after this step. 

\item
The VEV of the scalar field $\Delta_R$ or $\varphi_R$ breaks $SU(2)_L \otimes SU(2)_R \otimes  U(1)_{Y^{\prime \prime}}$ to the SM gauge group $SU(2)_L \otimes  U(1)_{Y^{\prime \prime \prime}}$.  After this step, the unbroken generator is $Y^{\prime \prime \prime} = T_{3R}+T^{\prime}_{3R} +T^{\prime}_{3L} + \frac{Y}{2}$

\item
Finally, the SM gauge symmetry   $SU(2)_L \otimes  U(1)_{Y^{\prime \prime \prime}}$ is broken down to the $U(1)_{EM}$ when the scalar field $\varphi$ acquires a VEV and we obtain electromagnetic charge operator $Q=T_{3L}+ T_{3R}+T^{\prime}_{3R} +T^{\prime}_{3L} + \frac{Y}{2}$.
\end{enumerate}

Thus, if we choose doublet representation to perform the SSB with the bi-doublets $\varphi$ and $\chi$, symmetry breaking pattern takes following form:

\begin{center}
\begin{gather}
\label{symmetrybreaking} 
	 SU(3)_c \otimes SU(2)_L \otimes SU(2)_R \otimes SU(2)^{\prime}_R \otimes SU(2)^{\prime}_L \otimes U(1)_{Y} 	\nonumber\\
		\downarrow \langle \chi_L \rangle 	\hspace*{0.1cm} ( \mathcal{W^\prime},~~ \mathcal{Z^\prime} )							\nonumber\\
SU(3)_c \otimes	SU(2)_L \otimes SU(2)_R \otimes SU(2)^{\prime}_R \otimes  U(1)_{Y^{\prime}}  				\nonumber\\
		\downarrow \langle \chi \rangle		\hspace*{0.1cm} ( \mathcal{X^\prime},~ \mathcal{Y^\prime} )													\nonumber\\
SU(3)_c \otimes	SU(2)_L \otimes SU(2)_R \otimes  U(1)_{Y^{\prime \prime}}															\nonumber\\
		\downarrow \langle \varphi_R \rangle   	\hspace*{0.1cm} ( \mathcal{X},~ \mathcal{Y} )												\nonumber\\
SU(3)_c \otimes	SU(2)_L \otimes  U(1)_{Y^{\prime \prime \prime}}      \nonumber\\
	\downarrow \langle \varphi \rangle  	\hspace*{0.1cm} ( \mathcal{W},~ \mathcal{Z} )												\nonumber\\
	SU(3)_c \otimes      U(1)_{EM},
\end{gather}
\end{center}
where $\mathcal{Y}$ is the neutral counter-part of the SM neutral gauge boson $\mathcal{Z}$. A possible pattern for the SSB is when VEVs are such that $\langle \chi_L \rangle  \geq \langle \chi \rangle >> \langle \varphi_R \rangle  > \langle \varphi \rangle$.  The detail study of the SSB is beyond the scope of this paper.

The invariance of the the gauge-scalar and the Yukawa sectors under parity dictates the behavior of the scalar Higgs fields under parity.  Thus, parity transformations of the Higgs fields are chosen as

\be
\label{sp1}
 \varphi  \longleftrightarrow \chi^\dagger,~ \Delta_L  \longleftrightarrow \Delta_R~ \textrm{or} ~\chi_L  \longleftrightarrow \varphi_R.
\ee

As a result, the scalar field $\varphi$  under $SU(2)_L \otimes SU(2)_R$, and scalar field  $\chi$ under $SU(2)^{\prime}_L \otimes SU(2)^{\prime}_R$ transform as following:

\be
\varphi \rightarrow U_L \varphi U^{\dagger}_R,~ \chi \rightarrow U^{\prime}_L \chi U^{\prime \dagger}_R.
\ee

The transformations of the scalar fields $\Delta_R$ and $\varphi_R$ under $SU(2)_L \otimes SU(2)_R $ are given by

\be 
\Delta_R \rightarrow U_R \Delta_R U^{\dagger}_R,~\varphi_R \rightarrow U_R \varphi_R .
\ee

The scalar fields $\Delta_L$ and $\varphi_L$ transform under $SU(2)^{\prime}_L \otimes SU(2)^{\prime}_R$ in the following way:
\be 
\Delta_L \rightarrow U^{\prime}_L \Delta_L U^{ \prime \dagger}_L,~\varphi_L \rightarrow U^{\prime}_L \varphi_L.
\ee

Now, we discuss the Yukawa sector of the model.  The most general Yukawa Lagrangian for quarks can be written 
\be
{\mathcal{L}}^{Q}_Y =  \bar{Q_L}  \left( \Gamma_1   \varphi + \Gamma_2 \tilde{\varphi} \right) Q_R + \bar{Q^{\prime}_{R}}  \left( \Gamma_1^{\prime \dagger}  \chi^\dagger + \Gamma_2^{\prime \dagger} \tilde{\chi}^\dagger \right) Q^{\prime}_{L} + h.c.,
\ee

where $\tilde{\varphi} = \tau_2 \varphi^* \tau_2$ and $\tilde{\chi} = \tau_2 \chi^* \tau_2$ are charge-conjugated Higgs fields and $\tau_2$ is the second Pauli-matrix. The $\Gamma_i$ and $\Gamma_i^\prime$ ($i=1,2$) are $3 \times 3$ matrices in generation space and  $\Gamma_i = \Gamma_i^{\prime \dagger}$ due to parity. 
 
The Yukawa Lagrangian for the leptons  can be written exactly in the same way.  The structure of the leptonic Yukawa Lagrangian will depend whether we use a doublet or a triplet representation of the Higgs fields to perform the SSB.  For example, if the doublets $\varphi_R$ and $\chi_L$ participate in the SSB along with the bi-doublets $\varphi$ and $\chi$, the Yukawa Lagrangian for the leptonic sector reads

\be
{\mathcal{L}}^{L}_Y =  \bar{L_L}  \left( \Pi_1   \varphi + \Pi_2 \tilde{\varphi} \right) L_R + \bar{L^{\prime}_{R}}  \left( \Pi_1^{\prime \dagger}   \chi^\dagger + \Pi_2^{\prime \dagger} \tilde{\chi}^\dagger \right) L^{\prime}_{L} + h.c..
\ee
Again, the $\Pi_i$ and $\Pi_i^\prime$ ($i=1,2$) are $3 \times 3$ matrices in generation space and  $\Pi_i = \Pi_i^{\prime \dagger}$ due to parity.

In summary, we have discussed a right-right-left  extension of the SM in this work.  The main features of the model, which are prime motivations to explore its phenomenological consequences are as follows:

\begin{itemize}
\item
The central idea of the present work is that the SM fermions live in the fundamental representations of the gauge groups $SU(2)_L $ and $ SU(2)_R $ and their coupling constants are completely independent.  This means, there is no parity invariance at this level.   Now, the question is whether parity can be restored. The one possible option is to embed $SU(2)_L $ and $ SU(2)_R $ in a larger symmetry.  The  other option is to assume that there are mirror symmetries $SU(2)^{\prime}_R$ and $SU(2)^{\prime}_L$ having mirror fermions in their fundamental representations.  This is the scenario in this model which is proposed.

\item
Since coupling constant of $SU(2)_L $ and $ SU(2)_R $ are independent,  this is a different scenario in contrast to the MLRSM where both coupling constant are same due to parity invariance.  Infact, it has been shown that a recent excess observed by the ATLAS and CMS collaborations\cite{Aad:2015owa,Aad:2014xka,Aad:2015ufa,Khachatryan:2014hpa,Khachatryan:2014gha,Khachatryan:2014dka,CMS:2015gla,Khachatryan:2015sja,Aad:2014aqa} may be explained with different coupling constant for $SU(2)_L $ and $ SU(2)_R $\cite{Dobrescu:2015qna,Dobrescu:2015yba,Deppisch:2014qpa,Deppisch:2014zta,Heikinheimo:2014tba,Aguilar-Saavedra:2014ola,Fowlie:2014mza,Krauss:2015nba,Gluza:2015goa,Dev:2015pga}.

\item
We also comment that recently observed di-photon excess\cite{ATLAS:di-photon,CMS:di-photon} by the ATLAS and CMS collaborations may be easily accommodated in this model with the help of new mirror fermions.

\item
There are no universal singlet fermions in the model under gauge groups $SU(2)$.  This is a unique feature of the model in compare to other models having mirror fermions \cite{Foot:1991bp,Silagadze:1995tr,Foot:1995pa,Berezhiani:1995yi,Gu:2012in,Chakdar:2013tca}.

\item
Parity restoration  occurs at the scale of  $SU(2)^{\prime}_L$.   This is again different from the MLRSM where parity is expected to restore at the scale of $SU(2)_R$.

\item
There is a four stage symmetry breaking in the model.  The MLRSM contains only a two stage symmetry breaking.

\item
One of the main and remarkable features is the scalar sector which is elegant and optimum.  There are four non-abelian gauge fields which acquire masses at the four stages of the SSB which occurs when exactly four scalar Higgs fields acquire the VEVs.  
\item
We note that the approach of this work is to extend gauge sector to accommodate new fermions.  On the other side, scalar sector has kept minimal.  Apart from $SU(2)_L $ and $SU(2)_R $, we have added two mirror symmetries $SU(2)_{R}^\prime$ and $SU(2)_{L}^\prime$ at the cost of only one scalar field $\chi$.

\item
The noted feature of the scalar sector is that the scale of the scalar Higgs fields $\Delta_L$ or $\chi_L$  is heavier than the scale of the Higgs fields $\Delta_R$ or $\chi_R$.  In the MLRSM, the scalar Higgs fields $\Delta_L$ or $\chi_L$ is supposed to have a tiny or vanishing VEV.  In this model, the VEV of $\Delta_L$ is the heaviest one.  This is a unique difference between the MLRSM and the proposed model in this work

\item
The symmetry product  $SU(2)_L \otimes SU(2)_R \otimes U(1)_{Y} $ is broken down to   $U(1)_{EM}$ when $\varphi$ and $\Delta_R$ acquire a VEV.  This feature is a main difference between this model and the MLRSM.  In the MLRSM, the same thing occurs when  $\varphi$, $\Delta_L$ and $\Delta_R$ acquire a VEV. 

\item
The bi-doublet $\varphi$ has a unique parity counter-part $\chi$.  In the MLRSM, the scalar field $\chi$ is absent.

\item
The scalar sector has a unique  pattern.  We have a light bi-doublet $\varphi$ and its heavy  parity counter part $\chi$.  There is a light triplet $\Delta_R$ and its heavy  parity counter part $\Delta_L$.

\item
There are two mirror symmetries $SU(2)_{R}^\prime$ and $SU(2)_{L}^\prime$ in the model which introduce six gauge bosons apart from the three gauge bosons associated with the group $SU(2)_{R}$.

\item
This is remarkable that gauge sector also has a unique pattern.  We have a light gauge boson $\mathcal{W}$ corresponding to the symmetry $SU(2)_L$ and a heavy gauge boson $\mathcal{W}^\prime$ associated with the group $SU(2)_{L}^\prime$. Similarly, there is a light gauge boson $\mathcal{X}$  corresponding to the symmetry $SU(2)_R$ and a heavy gauge boson $\mathcal{X}^\prime$ associated with the group $SU(2)_{R}^\prime$.  

\item
The model has a scope for spontaneous $CP$ violation.   The bi-doublets $\varphi$ and $\chi$ may have a complex VEV in general which could be the only source of $CP$ violation.  The mirror symmetries could have  a large $CP$ violation in comparison to the amount of the $CP$ violation observed in the SM and in turn, may provide a solution to the matter-antimatter asymmetry of the universe.
 
\item
We note that due to parity, Yukawa couplings and mass matrices of the SM quarks are Hermitian conjugate to those of their mirror counter-parts.  This means, at tree level, we obtain,

\be
\label{ph3}
\theta_{QFD} = -\theta^{\prime}_{QFD}.
\ee
where $\theta_{QFD}  = \textrm{arg~det} M_Q $ and similarly for $\theta^{\prime}_{QFD}$.  Thus, the strong $CP$ phase $\bar{\theta}^{\textrm{tree}} = 0$.

\end{itemize}

In the MLRSM,  the minimum scale of the gauge bosons corresponding to the gauge group $SU(2)_R$ with parity invariance is around 3 TeV \cite{Bertolini:2014sua}.  In the model proposed in this work, it may be possible to have the gauge bosons corresponding to the gauge group $SU(2)_R$ around 1-2 TeV.  

Now, we discuss the mass scale of the gauge bosons corresponding to  the gauge groups $SU(2)^{\prime}_R$,  $SU(2)^{\prime}_L$ and new mirror fermions.  In the mirror models discussed in the literature, for example in Refs. \refcite{Gu:2012in,Chakdar:2013tca}, the scale of gauge bosons corresponding to the gauge group $SU(2)_R$ is around $10^8$ GeV.  This is because the Yukawa couplings of the mirror fermions are identical to that of the SM ones due to parity.  Hence, if we wants to have heavy mirror fermions around 1 TeV, VEV of the Higgs field providing mass to gauge bosons corresponding to the gauge group $SU(2)_R$ should be of order $10^8$ GeV or so.  Thus, in compare to the MLRSM, the gauge bosons corresponding to the gauge group $SU(2)_R$ in these models are practically impossible to detect.  

In the model proposed, gauge bosons corresponding to the gauge group $SU(2)_R$ are within the reach of the LHC since the VEV of the triplet Higgs field $\Delta_R$ could be around a few TeV. This is an important advantage of the proposed model over other mirror models discussed in the literature \cite{Foot:1991bp,Silagadze:1995tr,Foot:1995pa,Berezhiani:1995yi,Gu:2012in,Chakdar:2013tca}.   However, since Yukawa couplings of the mirror fermions are identical to that of the SM ones due to parity, to keep masses of mirror fermions at TeV scale, the VEVs of the bi-doublet $\chi$  should be around $10^8$ GeV or so.  For example, for electron mass $0.511$ MeV, $v=\sqrt{v_1^2 + v_2^2} =246$ GeV and $v^\prime=\sqrt{v_1^{\prime2} + v_2^{\prime2}} = 10^8$ GeV, the mass of mirror electron is  208 GeV.  Hence, the masses of gauge bosons corresponding to the gauge groups $SU(2)^{\prime}_R$ and $SU(2)^{\prime}_L$ for $v^\prime= 10^8$ GeV   should be at least order of $10^8$ GeV.

The new mirror-fermions $\psi^{\prime}$  are heavy particles and are decoupled from the SM physics.  The neutrinos $\nu^\prime$ which are singlet under the SM could be a stable dark matter candidate. The quarks $u^{\prime}$ and $d^{\prime}$ have colour interactions under $SU(3)_c$ colour symmetry of the QCD.   Furthermore, charged mirror fermions also have electromagnetic-interactions due to the $U(1)_{EM}$ symmetry which is common among the SM and the mirror fermions.  Hence,  mirror fermions  can be  produced through strong as well as electromagnetic interactions. 

Since, mirror fermions and their SM counter-parts share same hypercharge, their electromagnetic charges are same.  Hence, as discussed in Ref. \refcite{Barr:1991qx}, there will be no fractionally charged hadrons and it should be possible to make mirror quarks light enough to satisfy baryon density of the universe.

In the non-minimal version of the model, for mirror fermions to decay in to the SM ones, we introduce a scalar bi-doublet $\eta$ which transforms as (2,2,2,2,0) under $SU(2)_L \otimes SU(2)_R \otimes SU(2)^{\prime}_R \otimes SU(2)^{\prime}_L \otimes U(1)_{Y}$.  Assuming that under parity, $\eta$ transforms as $ \eta \longleftrightarrow \eta^\dagger $, we write following Yukawa Lagrangian for the SM and mirror quarks with the bi-doublet $\eta$.

\be
{\mathcal{L}}_Y^\eta =  \bar{Q_L}  \left( \Sigma_1   \eta + \Sigma_2 \tilde{\eta} \right) Q_R + \bar{Q^{\prime}_{R}}  \left( \Sigma_1^{\prime \dagger}  \eta^\dagger + \Sigma_2^{\prime \dagger} \tilde{\eta}^\dagger \right) Q^{\prime}_{L} + h.c.,
\ee
where due to parity $\Sigma_i = \Sigma_i^{\prime \dagger}$ and they are $3 \times 3$ matrices in generation space.  A similar Lagrangian can be written for the leptonic sector.

The Yukawa Lagrangian leading to decays of the mirror quarks to the SM ones, is following:

\be
{\mathcal{L}}_Y =  \bar{Q_L}  \left( \mathcal{K}_1   \eta +  \mathcal{K}_2 \tilde{\eta} \right) Q_R^\prime +  \bar{Q_L}^\prime  \left( \Lambda_1   \eta +  \Lambda_2 \tilde{\eta} \right) Q_R   + h.c.,
\ee
where due to parity $\mathcal{K}_i = \mathcal{K}_i^{ \dagger}$ and $\Lambda_i = \Lambda_i^\dagger$. We can write a  similar Lagrangian for leptons.

The VEV of the Higgs bi-doublet $\eta$ lies at the electro-weak scale.  The Higgs bi-doublet $\eta$  introduces large flavour-changing neutral current on the SM side.  This can be cured by assumption of the alignment in flavour space which is proposed in Ref. \refcite{Pich:2009sp}. A similar approach is also proposed in Ref. \refcite{Joshipura:2010tz}.

Thus, we observe that the model contains interesting and rich phenomenological implications.  The model has bright prospects to test its many features in the future high luminosity running phase of the LHC.  The further details and phenomenological investigations will be provided in a future publication.

\section*{Acknowledgments}

I am grateful to Antonio Pich for the illuminating discussion during this work and many important suggestions on the manuscript.  Many thanks to  Anjan S. Joshipura, Namit Mahajan, Saurabh D. Rindani and Utpal Sarkar for valuable and critical feedback on this work.  I thank to Mehran Zahiri Abyaneh, Alessio Maiezza,  and Rahul Srivastava for various discussions.  This paper is dedicated to Aliza for her love, patience and everlasting support.  This work has been supported by the Spanish Government and ERDF funds from the EU Commission
[Grants No. FPA2011-23778, FPA2014-53631-C2-1-P No. and CSD2007-00042 (Consolider Project CPAN)].  

\section*{}

\end{document}